# Adversarial Attacks on Graph Neural Networks based Spatial Resource Management in P2P Wireless Communications

Ahmad Ghasemi, *Student Member, IEEE*, Ehsan Zeraatkar, Majid Moradikia, and Seyed (Reza) Zekavat, *Senior Member, IEEE*

*Abstract*—This paper introduces adversarial attacks targeting a Graph Neural Network (GNN)-based radio resource management system in point-to-point (P2P) communications. Our focus lies on perturbing the trained GNN model during the test phase, specifically targeting its vertices and edges. To achieve this, four distinct adversarial attacks are proposed, each accounting for different constraints, aiming to manipulate the behavior of the system. The proposed adversarial attacks are formulated as optimization problems, aiming to minimize the system's communication quality. The efficacy of these attacks is investigated against the number of users, signal-to-noise ratio (SNR), and adversary power budget. Furthermore, we address the detection of such attacks from the perspective of the Central Processing Unit (CPU) of the system. To this end, we formulate an optimization problem that involves analyzing the distribution of channel eigenvalues before and after the attacks are applied. This formulation results in a Min-Max optimization problem, allowing us to detect the presence of attacks. Through extensive simulations, we observe that in the absence of adversarial attacks, the eigenvalues conform to Johnson's SU distribution. However, the attacks significantly alter the characteristics of the eigenvalue distribution, and in the most effective attack, they even change the type of the eigenvalue distribution.

## I. INTRODUCTION

THANKS to the powerful capability of Machine Learning (ML) in solving extremely complex communications problems, such as radio resource management, [2], [3], beam prediction [4], channel estimation [5], localization [6]–[8], and spectrum sensing [9], they have attracted a great deal of interest in academia and industry. However, ML, and in particular deep learning (DL) algorithms suffer from: 1) Poor generalization as they require a large amount of data for training, and, 2) poor scalability since their performance decreases as the problem size increases. To tackle these problems, graph neural networks (GNNs) have been recently introduced by integrating graph theory and DL. GNNs tackle these problems using different approaches such as graph-wise sampling, which constructs sub-graphs for the model inference [10]. GNN algorithms provide powerful tools that has shown great success in many domains such as Computer Vision (CV) [11], [12] and Natural Language Processing (NLP) [13], [14]. Motivated by the success of GNNs in these domains, GNNs have been recently used in wireless communications [2], [3], [15]–[24].

However, similar to other ML algorithms, GNN-based wireless communications have been found vulnerable to manipulations and perturbations generated by adversaries that may target the training and/or testing process of DL [25], [26]. The corresponding study field, known as adversarial machine learning (AML), focuses on developing adversarial attack algorithms or defense algorithms to secure DL systems. One common form of AML called adversarial attack occurs when an adversary crafts small perturbations to the original input of a neural network or develops a means to manipulate the operation of a DL system to cause an error in the inference process. These perturbations are more destructive compared to jamming attacks [25] and are carefully crafted to produce a vector in the input feature space that is capable of misleading the DL algorithm. Therefore, it is also different from spoofing attacks, where the adversary generally aims to impersonate a legitimate user or device, sending false information or messages that alter the perceived feedback.

Such adversarial attacks pose a significant threat to modern wireless network systems, including point-to-point (P2P) wireless communications such as device-to-device (D2D), machine-to-machine (M2M), and vehicle-to-vehicle (V2V) communication [27]. These systems are crucial components of modern wireless networks, facilitating efficient data transfer between short-range mobile devices. P2P wireless communications find applications in military systems as well as in 5G mobile communications to improve spectral efficiency [28]. For instance, D2D enables local data services via unicast transmission and can extend coverage to assist users at the cell edge or in disaster-hit areas. Moreover, it supports M2M communication, a key technology for the Internet of Things (IoT), involving autonomous connectivity and communication among IoT devices [29], [30].

Adversarial attacks targeting GNN-based radio resource management systems have significant implications for various real-world applications. In the realm of IoT networks, where efficient communication protocols are crucial for the increasing prevalence of IoT devices, these attacks pose a threat. An imperfect resource allocation in IoT networks leads to performance degradation, increased latency, and potential

A. Ghasemi, M. Moradikia, and S. A. Zekavat are with the Department of Data Science, Worcester Polytechnic Institute (WPI), 100 Institute Road, Worcester, MA, 01609-2280, USA, e-mail: {aghasemi2, mmoradikia, rezaz}@wpi.edu., E. Zeraatkar is with Shiraz Urban Railway Organization, Shiraz, Iran, email: ezeraatkar@gmail.com.

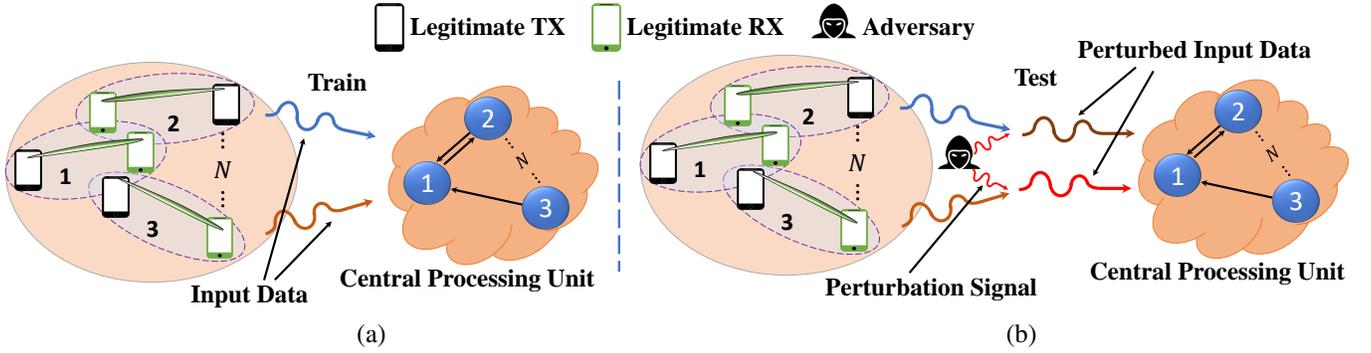

Fig. 1: System model: (a) training phase, (b) testing phase with Evasion attack.

security risks [31]. In addition, Vehicular ad hoc networks (VANETs) that facilitate communications between vehicles and roadside infrastructure in intelligent transportation systems are subject to adversarial attacks. These attacks can compromise the communication quality in VANETs, affecting traffic flow, road safety, and overall transportation system efficiency [32]. Moreover, in smart grid, efficient communication among different components is vital for reliable power distribution. However, adversarial attacks on GNN-based resource management systems in smart grids can disrupt the process, resulting in unreliable power distribution and potential operational risks [33]. Accordingly, we need to investigate adversarial attacks to facilitate secure and efficient operation of many real-world applications.

Against this background, to the best of our knowledge, no work has evaluated adversarial attacks against GNN-based P2P wireless communications so far. Although security issues, such as identifying and defending against adversarial attacks, are crucial for wireless communications, we must first put ourselves in the place of adversaries to quantify their destructive effects. Specifically, this paper focuses on proposing practical adversarial attacks on GNN-based P2P wireless communications and detecting their occurrence.

Against the background described earlier, our contributions are summarized as follows:

- In the first step, we consider adversarial attacks against GNN-based P2P wireless communications. Specifically, we propose four new adversarial attacks, two of which target vertices of the trained GNN model and the other two deteriorate the edges during the testing phase. We consider three constraints in the adversarial attack design: 1. Channel-bounded [34], [35]: adversaries are constrained by the number of perturbations that they can make simultaneously, 2. Power-bounded [36]: adversaries have limited power, and 3. Min-detectable: adversaries aim to perturb the information in a way that minimizes the probability of being detected by the system. We have formulated the adversarial problem in a way that is agnostic of the specific deep learning architecture, indeed broadens the scope of the proposed approach, making it potentially applicable to a variety of network controllers.
- To detect those attacks that are carefully designed so that they cannot be detected by the system's defense mechanism, we propose a new Min Max optimization problem based on the *One-Sample Kolmogorov-Smirnov (KS) Test*. By doing so, we will be able to identify the cumulative distribution function (CDF) of eigenvalues of the channel tensor (including the entire channels between the legitimate TXs and RXs) before and after applying the adversarial attacks. This optimization problem searches among 84 different theoretical CDFs and finds the one having the minimum value of the largest difference (in absolute value) between the observed and theoretical CDFs. The results show that the eigenvalues follow Johnson's SU distribution [37] when there is no adversarial attack. In addition, our results validate that adversarial attacks change the properties of the CDF of eigenvalues, and particularly the type of CDF in the most effective proposed adversarial attack.
- Considering the limited resource budget available, the adversary on the other hand tries to solve another optimization problem with the aim of maximizing the destructive impact on the total quality of communication (QoC), which is a weighted sum rate, of the considered network. The results illustrate the effectiveness of the proposed adversarial attacks as they could result in a 95% decrease in the total QoC.

### A. Organization and Notation

The rest of the paper is organized as follows: Section III describes the system model and problem definition. The proposed adversarial attacks are introduced in Section IV, while Section V determines the distribution of channel tensor eigenvalues. The proposed approaches are evaluated in Section VI. Finally, Section VII concludes the paper.

**Notation:** In this paper, vectors are denoted by small bold-italic face letters $\boldsymbol{a}$, and capital bold-italic face letters $\boldsymbol{A}$ represent matrices. The rank of matrix $\boldsymbol{A}$ is denoted by $\text{rank}(\boldsymbol{A})$. $\mathcal{A}$ is a set and $a$ is a scalar. The $i^{\text{th}}$ element and the number of elements of set $\mathcal{A}$ or the cardinality of this set, are shown via $\mathcal{A}[i]$ and $|\mathcal{A}|$, respectively. $|a|$ and $\angle a$ are the magnitude and phase of the complex number $a$. We also define two new element wise operators for vectors: $|\dot{\boldsymbol{a}}| \triangleq [|a_0|, |a_1|, ..., |a_{N-1}|]^{\text{T}}$ and $\dot{\angle}\boldsymbol{a} \triangleq [\angle a_0, \angle a_1, ..., \angle a_{N-1}]^{\text{T}}$, where $a_i, \forall\ i = 0, ..., N-1$, are the elements of the vector $\boldsymbol{a}$. The transpose and Hermitian (conjugate transpose) of a matrix/vector are shown by $(.)^{\mathrm{T}}$, $(.)^{\dagger}$, respectively. $\|.\|_2$ denotes $l_2$-norm of a vector. $\mathbb{D}^{l \times l}$ and $\mathbb{C}^{m \times n}$ represent a diagonal matrix of dimension $l \times l$ and a complex matrix of dimension $m \times n$. The $n^{\text{th}}$ diagonal element of a diagonal matrix $\mathbb{D}$ is denoted by $\mathbb{D}_n$. $\mathbb{R}$ denotes the set of all real numbers. $\mathbf{I}_N$ denotes the $N \times N$ identity matrix. $\mathbf{0}_N$ and $\mathbf{1}_N$ are the $N$-dimensional all-zeros and all-ones vectors, respectively. We use $\mathcal{CN}(\mu, \sigma^2)$ to denote a circularly symmetric complex Gaussian random vector with mean $\mu$ and variance $\sigma^2$. Finally, $P(.)$, $(.)^*$, and $\mathbb{E}(.)$ denote the probability, the optimum value, and the expectation, respectively.

## II. Preliminaries

### A. Graph Neural Network

GNNs are a set of deep learning methods that operate in the graph domain. A graph is a data structure composed of edges and vertices/nodes. The relationships between the various vertices are defined by the edges. If the direction of the edges is specified, the graph is called directed; otherwise, it is called undirected. Unlike other types of data such as images, graph data operates in a non-Euclidean space. GNNs are neural networks that can be directly applied to graphs, providing a straightforward approach to predicting vertex-level, edge-level, and graph-level outputs. Vertex-level and edge-level outputs require a prediction for each vertex and edge in a given graph, respectively. Additionally, a graph-level output requires a prediction for the graph as a whole, which is the focus of this paper.

### B. Adversarial Attack

An adversarial attack (also known as an adversarial example) involves feeding a carefully crafted input signal to an ML model in order to cause it to fail or produce false outputs. Adversaries often keep the corresponding signal power below a threshold to conceal their presence in the network. While recent advances in adversarial attacks for text and images have been made, such as AddAny, AddSent [38], and FGSM [39], only a few works have incorporated graph structures. The combinatorial nature of graph structures makes the problem much more difficult compared to text, and unlike images, graph data is discrete, making it a non-trivial problem to effectively attack graph structures. Nonetheless, there have been works in adversarial attacks against graph-structured data [40]–[46].

There are three well-known types of adversarial attacks based on the amount and type of available information at the adversary's disposal [25], [47]: 1) White-box, where all information such as exact channel information and DL model is known to the adversary, 2) Grey-box, where limited information such as the channel distribution is available, and 3) Black-box, where no information is available to the adversary. From the perspective of training and testing phases, adversarial attacks can be classified into two general schemes [48]: 1) *Poisoning*, where the adversary affects the performance of a DL model by injecting adversarial samples into the training dataset during the training phase, and 2) *Evasion*, where the parameters of the trained model are fixed, and the adversary changes the input data during the testing phase.

### C. Adversarial Attacks against ML/DL-based Wireless Communications

There are several studies in the literature exploring the application of adversarial attacks in wireless communications [25], [26], [49]–[54]. In [25], the authors investigated various adversarial attacks against an autoencoder communication system. They demonstrated that an adversary can cause significant disruption even without perfect knowledge of the DL-model of the autoencoder system and without synchronicity with the transmitter. Adversarial attacks against power allocation scenarios were considered in [26], where a BS allocates transmit power to orthogonal subcarriers using a deep neural network (DNN). The input and output of the DNN are channel gains and transmit powers, respectively. An adversary perturbs the input data, which consists of channel estimations, to the DNN in order to decrease the system's sum-rate. In [49], an evasion attack is applied by adding small perturbations to the original data samples to create adversarial examples that result in misclassification at the receiver. The adversary in [51] has direct access to the input of a modulation classifier, which is transmitted over the air (OTA) or asynchronously from a separate device. In [53], an adversary perturbs channel input symbols at the encoder, resulting in the encoder incorrectly classifying modulations with a high level of confidence. Furthermore, an over-the-air attack is introduced in [54] to manipulate a reinforcement learning (RL) algorithm. This attack is used to interfere with network slicing by manipulating the RL algorithm. The attacker is able to observe the spectrum and construct their own RL model, which is used to select resource blocks to jam while adhering to an energy budget. The objective of the attack is to maximize the number of failed requests caused by the jammed resource blocks.

It is worth noting that none of the above-mentioned works are specifically focused on graph-structured models and data, and therefore, they are not directly applicable to the scenario considered in our problem. Although there are several works that apply adversarial attacks on graph-based systems - as mentioned in the previous subsection - they are not within the domain of wireless communications and are not directly applicable to P2P communications. This is mainly because they assume availability of either the learning model or its gradients, which is not easily satisfied in wireless communications. In this work, we assume that this information is not available to the adversary, and the adversary only has partial information about the input data to the model. Furthermore, none of the above-mentioned works specifically focus on spatial resource management, which is the main focus of our paper.

## III. System model and problem definition

We are considering a multi-user multi-input single-output (MISO) wireless network, which comprises $N$ active transceiver pairs in the set of $\mathcal{N} \in \{1, 2, ..., N\}$, where TXs incorporate $N_t$ antenna elements and RXs are single antenna devices (Fig. 1 (a)). Let $\{s_n\}_{n=1}^{N}$ be the unit-norm transmitted signal from the $n^{\text{th}}$ TX to the $n^{\text{th}}$ RX, and let $\mathbf{Q} = [\mathbf{q}_1, \mathbf{q}_2, ..., \mathbf{q}_N]^{\mathrm{T}} \in \mathbb{C}^{N \times N_t}$, with $\{\mathbf{q}_n\}_{n=1}^{N}$ denoting the $n^\text{th}$ transmitter precoder (TPC), the estimated received signal at the $n^\text{th}$ RX is formulated as $y_n = \mathbf{h}_{n,n}^\dagger \mathbf{q}_n s_n + \sum_{i=1,i\neq n}^{N} \mathbf{h}_{i,n}^\dagger \mathbf{q}_i s_i + n_n$. Here, $\mathbf{h}_{i,n} \in \mathbb{C}^{N_t}$ denotes the channel vector from the $i^\text{th}$ TX to the $n^\text{th}$ RX, and $n_n \sim \mathcal{CN}(0, \sigma_n^2)$ represents the Additive White Gaussian Noise (AWGN) at the $n^\text{th}$ RX.

Furthermore, all channels can be included in a channel tensor $\mathbf{H} \in \mathbb{C}^{|\mathcal{V}|\times|\mathcal{V}|\times N_t}$. Elements of this channel tensor are $\mathbf{H}_{i,n,:} = \mathbf{h}_{i,n} \in \mathbb{C}^{N_t}, \{i,n\} \in \mathcal{N}$, where diagonal and non-diagonal elements respectively correspond to the desired and interference channels of transceiver pairs. This channel tensor is known to both the central processing unit (CPU) and the adversary. The CPU has the role of building and updating the DL model.

### A. Graph modeling of P2P Wireless Communications

As shown in Fig. 1, a *directed graph* is used to model the considered P2P wireless network, where the $n^\text{th}$ transceiver pair represents the $n^\text{th}$ vertex in the graph. The vertex features encompass the properties of the transceivers. A directed edge from a vertex $i$ towards a vertex $j$ is depicted if interference from the vertex $i$ as TX is imposed on the vertex $j$ as RX and the edge feature includes the properties of the corresponding interference channel between TX $i$ and RX $j$. Notably, the interference exists only if the distance between TX $i$ and RX $j$ is below a certain threshold $T_d$. Formally, a graph $\mathcal{G}(\mathcal{V}, \mathcal{E})$ represents the graph model of P2P wireless communications, where $\mathcal{V}$ and $\mathcal{E}$ are the set of vertices and edges, respectively. In this graph, the vertex feature matrix $\mathbf{Z} \in \mathbb{C}^{|\mathcal{V}|\times(N_t+2)}$ is given by $\mathbf{Z}_{n,:} = [\mathbf{h}_{n,n}, w_n, \sigma_n^2]^\text{T}$, in which $|\mathcal{V}|$ stands for the cardinality of the vertex set $\mathcal{V}$ and $w_n$ represents the weight for the $n^\text{th}$ pair or vertex. Here, the $(N_t+2)$ dimension arises from the intrinsic properties of the system under consideration. Specifically, the length of $\mathbf{h}_{n,n}$ is $N_t$, representing the number of transmitter antennas. The additional '2' accounts for the variables $w_n$ and $\sigma_n^2$, each being scalar quantities. Thus, the resulting dimension of the complex space $\mathbf{Z}$ is $\mathbb{C}^{|\mathcal{V}|\times(N_t+2)}$, embodying all the relevant system parameters. Moreover, the adjacency feature tensor $\mathbf{A} \in \mathbb{C}^{|\mathcal{V}|\times|\mathcal{V}|\times N_t}$ is given by:

$$\mathbf{A}_{i,n,:} = \begin{cases} \mathbf{0}_{N_t}, & \text{if } \{i,n\} \notin \mathcal{E}, \\ \mathbf{h}_{i,n}, & \text{otherwise}, \end{cases} \quad (1)$$

where, $\mathbf{h}_{i,n} \in \mathbb{C}^{N_t}$, $\{i,n\} \in \mathcal{N}$, represents the $N_t$-length channel vector from the $i^\text{th}$ TX to the $n^\text{th}$ RX. Now using the definitions of variables $\mathbf{Z}$ and $\mathbf{A}$, the received signal at the $n^\text{th}$ RX can be reformulated as:

$$y_n = \underbrace{\mathbf{Z}_{n,1:N_t}^\dagger \mathbf{q}_n s_n}_{\text{Desired Signal}} + \underbrace{\sum_{\substack{i=0 \\ i\neq n}}^{N} \mathbf{A}_{i,n,:}^\dagger \mathbf{q}_i s_i}_{\text{Interference}} + \underbrace{n_n}_{\text{Noise}}, \quad (2)$$

thus, the Signal-to-Interference-plus-Noise Ratio (SINR) at the $n^\text{th}, n \in \mathcal{N}$, RX of the $n^\text{th}, n \in \mathcal{N}$ vertex (transceiver pair) is determined in (3), where, $\mathbf{Z}_{n,N_t+2}$ denotes the noise power, which is determined according to the definition of the vertex feature matrix $\mathbf{Z}$.

$$\text{SINR}_n = \frac{|\mathbf{Z}_{n,1:N_t}^\dagger \mathbf{q}_n|^2}{\sum_{i=1,i\neq n}^{N} |\mathbf{A}_{i,n,:}^\dagger \mathbf{q}_i|^2 + \mathbf{Z}_{n,N_t+2}}, \quad (3)$$

To train the considered graph-based wireless communication system, the GNN model of [15] is used, which includes three layers. The channel states $\{\mathbf{Z}_{n,1:N_t}^\dagger\}_{n=1}^{N}$ and users' weights $\{w_n\}_{n=1}^{N}$ are used as the input of the GNN. The CPU uses all the information received from the transceivers to train a centralized GNN-model of the system (Fig. 1 (b)). At the output, the GNN yields the beamforming vectors of users by minimizing the loss function $l_\Theta$ (4) at the last layer. Here, $\Theta$ represents the weights of GNN, which are trained in an adversarial-free environment at the CPU. Thus, by considering the average over all generated channels, the CPU aims to minimize $l_\Theta$, which is given by:

$$l_\Theta = -\mathbb{E}\Big(\sum_{n=1}^{N} \mathbf{Z}_{n,N_t+1} \log_2(1 + \text{SINR}_n(\Theta))\Big), \quad (4)$$

where, $\mathbf{Z}_{n,N_t+1}$ represents the weights of users as per the established definition of $\mathbf{Z}$ in our study, and $\text{SINR}_n(\Theta)$ is defined as:

$$\text{SINR}_n(\Theta) = \frac{|\mathbf{Z}_{n,1:N_t}^\dagger \mathbf{q}_n(\Theta)|^2}{\sum_{\substack{i=1 \\ i\neq n}}^{N} |\mathbf{A}_{i,n,:}^\dagger \mathbf{q}_i(\Theta)|^2 + \mathbf{Z}_{n,N_t+2}}, \quad (5)$$

where, $\mathbf{Z}_{n,N_t+2}$ is referenced back to its initial definition provided after equation (3).

## IV. ADVERSARIAL ATTACKS

In this section, we first outline the important assumptions about the adversary in this paper.

*Remark.* **(Adversary's Important Presumptions):** 1) Adversary is a white-box adversary that has access only to channel information; 2) Adversary performs evasion attacks; 3) Adversary incorporates multiple antenna elements to transmit perturbations and attack multiple channels simultaneously, as each transceiver pair sends its information to the CPU through different channels; Using a single antenna, it can only attack one channel at a time; 4) Adversary limits its perturbation's power to reduce the likelihood of attack detection by the CPU and legitimate users. 5) All transceiver pairs are within the transmission range of the adversary [34], [55]; 6) Adversary can listen and learn transceiver pairs' information; 7) Adversary can choose different channels, freq. and so on to disrupt at any fraction of time; 8) Adversary is a reactive adversary, meaning that it performs physical carrier sensing (which is a part, e.g., of the 802.11 standard) to learn whether the channel is currently idle or busy; 9) Adversary can introduce malicious messages using the address spoofing [34], [55]. ∎

In our system model, the adversary actively participates in the P2P network and has control over the channels transmitted to the CPU. During the testing phase, the adversary's objective is to degrade the performance of the trained network by minimizing the QoC of transceiver pairs. This QoC is measured as the weighted sum rate, denoted as $\sum_{n=1}^{N} \mathbf{Z}_{n,N_t+1} \log_2(1 + \text{SINR}_n)$, where $N$ is the number of transceiver pairs, $\mathbf{Z}_{n,N_t+1}$ represents a weight associated with each pair, and $\text{SINR}_n$ is the SINR pair $n$ [56]. To achieve this objective, the adversary interferes with the channel information, $\mathbf{h}_{i,n}$, transmitted by each transceiver pair to the CPU. Transceiver pairs initially determine these channel information values and transmit them to the CPU. The CPU relies on this information to train the GNN model. However, during the testing phase, the adversary manipulates the information transmitted by the transceiver pairs, resulting in perturbed channel information being received by the CPU. We will see that this deliberate interference by the adversary significantly impacts the network's performance.

Note that we use the terms "channel" and "channel information" interchangeably, as both represent the information of the transceiver pairs towards the CPU. The adversary needs to first acquire this knowledge, i.e., $\mathbf{h}_{i,n}$. There are two possible approaches available for the adversary to obtain $\mathbf{h}_{i,n}$: 1) It can impersonate itself as a fake CPU for a shot time-frame and trick users into sending data to it. By doing so, even when the real CPU is available, the adversary will be able to inject malicious packets into users to get the required info from [57], 2) It can persistently observe the data to learn from it. By doing so, it will be able to extract knowledge about the pairs and their corresponding channel information. By exploiting the extracted information together with power budget constraints over the network, the adversary now searches for the values of pairs channel information $\widehat{\mathbf{h}}_{i,n}$ to minimize the system's QoC. The objective of the adversary is to minimize the total Quality of Communication (QoC) by perturbing the channel information of the vertices in the set $\mathcal{A}$. Accordingly, the corresponding general optimization problem is defined as:

$$\min_{\widehat{\mathbf{h}}_{i,n}} \quad \sum_{n \in \mathcal{A}} \mathbf{Z}_{n,N_t+1} \log_2(1 + \widehat{\text{SINR}}_n), \tag{6a}$$

$$\text{s.t.} \quad \mathcal{A} \subseteq \mathcal{N} \text{ and } |\mathcal{A}| \geq 1 \tag{6b}$$

$$\|\mathbf{q}_n\|_2^2 \leq P_{\max}, \ \forall \ n \in \mathcal{N} \tag{6c}$$

where, $\widehat{\text{SINR}}_n$ denotes the perturbed version of $\text{SINR}_n$ in (3), in which the perturbed channel information $\widehat{\mathbf{h}}_{i,n}$ are replaced. In addition, $\mathcal{A}$ denotes the set of vertices targeted by the adversary for attack, and the corresponding constraint (6b) enforces that the set $\mathcal{A}$ contains at least one vertex and is a subset of the total vertex set $\mathcal{N}$. It explicitly defines the characteristics and limitations of the set $\mathcal{A}$ chosen by the adversary. Moreover, the constraint (6c) expresses the power budget constraint at TXs.

*Remark.* (**Adversary's Constraints**): Two constraints are typically considered for adversaries in wireless communication [34]–[36]: L1) **Channel-bounded ($B_c$)**: In this type, the number of channels that an adversary can attack simultaneously is limited; This limitation will be considered for designing $B_c$ perturbations in Subsection IV-A; and L2) **Power-bounded ($B_p$)**: In this type, the adversary has a limited power budget. This constraint, together with L1, is considered for designing $B_p$ perturbations in Subsection IV-B. ∎

*Remark.* (**Adversarial Attacks in Graphs**): Adversaries can attack vertices and/or edges of the graph to change their respective features. This paper considers both types: R1) Adversaries who are able to attack the set of vertices $\mathcal{V}$ of the graph $\mathcal{G}$ where they change desired channels between TXs and RXs. R2) Adversaries who can attack the set of edges $\mathcal{E}$ of $\mathcal{G}$ in which they manipulate interference channels. ∎

Based on the combination of L1 and L2 with R1 and R2, described above, four different adversarial attacks are introduced in the sequel. Specifically, $\widehat{\text{SINR}}_n$ will be defined based on the type of adversarial attack. Additionally, in each proposed attack, a set of constraints is added to (6).

In the proposed attacks, the adversary transmits a low power perturbation signal $s_p$ is designed as a function of channel information. As such, the CPU will receive $x_{\text{adv}} = x + s_p$, where, $x$ is the clean information transmitted by the transceivers, and $x_{\text{adv}}$ is the perturbed information received at the CPU. The goal of the adversary is to design $s_p$ in such a way that it causes the DL model at the CPU to receive misinformation during the testing phase, without being detected.

### A. $B_c$ Perturbations

This subsection introduces $B_c$ vertex and $B_c$ edge perturbations by considering L1.

*1) $B_c$ vertex perturbation:* In this attack, the adversary perturbs the channel information of transceiver pairs or vertices of the graph $\mathcal{G}$ to degrade the SINR defined in (3). These channel coefficients constitute the diagonal elements of $\mathbf{H}$, i.e., $\{\mathbf{h}_{i,i} \triangleq [h_{i,i,1}, h_{i,i,2}, ..., h_{i,i,N_t}]\}_{i=1}^{N} \in \{\mathbf{H}_{n,n,:}\}_{n=1}^{N}$ or, equivalently, the first $N_t$ rows of the vertex feature matrix $\mathbf{Z}$. Given $\mathbf{H}$ as well as $0 \leq l_c \leq 1$, which determines the percentage of perturbed channel information towards the CPU, the adversary randomly selects the victim transceiver pair's channel information included in the vertex set $\mathcal{V}$. Observing (7), the objective function (OF) strictly increases with respect to $\{\widehat{\mathbf{H}}_{n,n,:}\}_{n=1}^{N}$ because $\{\widehat{\mathbf{h}}_{i,i} \triangleq [\widehat{h}_{i,i,1}, \widehat{h}_{i,i,2}, ..., \widehat{h}_{i,i,N_t}]\}_{i=1}^{N_a} \in \{\widehat{\mathbf{H}}_{n,n,:}\}_{n=1}^{N}$ appear in the *numerator* of the OF. Thus, to reach the minimum SINR, the perturbation signal $\mathbf{s}_p$ must be crafted so that $\{\widehat{\mathbf{H}}_{n,n,:}\}_{n=1}^{N}$ are *reduced*. In addition, to prevent detection, the adversary aims to keep these channel gains higher than a threshold $h_{\text{th}}$ determined by the CPU. The CPU indicates this value based on its observations of the channels' history. If the received channel gains at the CPU are less than this value, the CPU notices that there is some abnormality in the system and requests TXs to resend their data. If this problem occurs several times, the CPU can detect the occurrence of the attack. In practice, $h_{\text{th}}$ is unknown to the adversary. Hence, during the perturbation, the adversary determines the minimum value of the channel gains of vertices denoted by $h_{\text{diag,min}}$ and tries to keep the gains of the perturbed channels above this new threshold as stipulated in (7d). Note that by doing this, the detection probability decreases significantly but does not go to zero because this new threshold does not necessarily match to $h_{\text{th}}$. As discussed above, the corresponding optimization problem for the $B_c$ vertex perturbation is formulated in (7), which includes two additional constraints (7d) and (7e) compared to the original

$$\min_{\widehat{\mathbf{H}}} \quad \sum_{n \in \mathcal{A}} \mathbf{Z}_{n,N_t+1} \log_2(1 + \widehat{\text{SINR}}_n) = \sum_{n \in \mathcal{A}} \mathbf{Z}_{n,N_t+1} \log_2(1 + \frac{|\widehat{\mathbf{H}}_{n,n,:}^{\dagger} \mathbf{q}_n|^2}{\sum_{\substack{i=1 \\ i \neq n}}^{N} |\mathbf{H}_{i,n,:}^{\dagger} \mathbf{q}_i|^2 + \mathbf{Z}_{n,N_t+2}}) \quad (7a)$$

$$\text{s.t.} \quad C_0: \quad \mathcal{A} \subseteq \mathcal{N} \quad (7b)$$

$$C_1: \quad \|\mathbf{q}_i\|_2^2 \leq P_{\max}, \ \forall \ i \in \mathcal{N} \quad (7c)$$

$$C_2: \quad |\widehat{\mathbf{h}}_{i,i}| \geq h_{\text{diag,min}} \mathbf{1}_{N_t}, \ \forall \ \{\widehat{\mathbf{h}}_{i,i} \triangleq [\widehat{h}_{i,i,1}, \widehat{h}_{i,i,2}, ..., \widehat{h}_{i,i,N_t}]\}_{i=1}^{N_a} \in \{\widehat{\mathbf{H}}_{n,n,:}\}_{n=1}^{N} \quad (7d)$$

$$C_3: \quad 1 \leq N_a \leq \min(L, N) \quad (7e)$$

---

optimization problem in (6). Remarkably, the constraint (7e) arises due to the limitation on the maximum number of perturbed channels, $N_a = |\mathcal{A}|$, which is determined by the minimum value between I) the total number of transceiver pairs, $N$, and II) the number of antenna elements available to the adversary, $L$.

It has been shown in [15] that the original problem (6) is non-convex and thus difficult to solve. The situation is exacerbated in (7) by adding constraints (7d) and (7e). To tackle the resulting non-convexity, we propose a *heuristic* algorithm to achieve a suboptimal but reasonably good solution. The main part of this algorithm is the perturbation signal design. It consists of two parts: one part nullifies the channel information of randomly selected transceiver pairs to minimize (7) as much as possible without considering (7d), and the second part satisfies (7d). Assume the corresponding perturbation signal for the randomly selected $\mathbf{h}_{i,i}$ is $\mathbf{s}_{p,i}$. For the first part, we intuitively conclude that the perturbed channel $\widehat{\mathbf{h}}_{i,i}$ must have an analogous structure to $\mathbf{h}_{i,i}$ but with a $\pi$ radian phase difference, i.e., $\widehat{\mathbf{h}}_{i,i} \leftarrow |\mathbf{h}_{i,i}| e^{j(\angle \mathbf{h}_{i,i} + \Pi)}$, where $\Pi = \pi \mathbf{1}_{N_t}$, so that the resultant SINR in (5) of the $i^{\text{th}}$ transceiver pair will be zero at the CPU. For the second part, as mentioned earlier, to satisfy the minimum value for the channel gain and thus avoid detection by the CPU, $s_{p,i}$ includes an additional part with the same phase as the original channel but with an amplitude equal to $h_{\text{diag,min}}$, i.e., $h_{\text{diag,min}} e^{j \angle \mathbf{h}_{i,i}}$. Thus, the perturbation signal is $\mathbf{s}_{p,i} \leftarrow \widehat{\mathbf{h}}_{i,i} + h_{\text{diag,min}} e^{j \angle \mathbf{h}_{i,i}}$. The $B_c$ vertex perturbation is summarized in **Algorithm 1**. As a result of this perturbation, the values of $\{\text{SINR}_i(\Theta)\}_{i=1}^{N_a}$ computed at the CPU will have very small values.

---

**Algorithm 1** $B_c$ vertex perturbation

**Input:** $\mathbf{H}$, $l_c$
1: Obtain $\mathcal{H}_{\text{diag}} := \{\mathbf{H}_{n,n,:}\}_{n=1}^{N}$ and $h_{\text{diag,min}} = \min \mathcal{H}_{\text{diag}}$,
2: $N_a = \lfloor N \times l_c \rfloor$,
3: **for** $i = 1, 2, ..., N_a$ **do**
4:     Randomly select one element of $\mathcal{H}_{\text{diag}}$ as $\mathbf{h}_{i,i} = |\mathbf{h}_{i,i}| e^{j \angle \mathbf{h}_{i,i}}$
5:     $\widehat{\mathbf{h}}_{i,i} \leftarrow |\mathbf{h}_{i,i}| e^{j(\angle \mathbf{h}_{i,i} + \Pi)}$
6:     $\mathbf{s}_{p,i} \leftarrow \widehat{\mathbf{h}}_{i,i} + h_{\text{diag,min}} e^{j \angle \mathbf{h}_{i,i}}$
7:     $\mathbf{h}_{i,i} \leftarrow \mathbf{h}_{i,i} + \mathbf{s}_{p,i}$
8: $\mathbf{H} \leftarrow$ Update $\mathbf{H}$ with the modified $\mathcal{H}_{\text{diag}}$
**Output:** $\widehat{\mathbf{H}}$,

---

*2) $B_c$ edge perturbation:* In this attack, the adversary perturbs the channel information of interference channels or edges of the graph $\mathcal{G}$ to degrade the SINR defined in (3). These channel coefficients form the non-diagonal elements of $\mathbf{H}$, i.e., $\{\mathbf{h}_{i,n} \triangleq [h_{i,n,1}, h_{i,n,2}, ..., h_{i,n,N_t}]\}_{i,n=1, i \neq n}^{N} \in \{\mathbf{H}_{i,n,:}\}_{i \neq n, \{i,n\} \in \mathcal{N}}$ and non-zero elements of the adjacency feature matrix $\mathbf{A}$ in (1). The OF in (8) is strictly decreasing with respect to $\{\widehat{\mathbf{H}}_{i,n,:}\}_{i,n=1, i \neq n}^{N}$ because $\{\widehat{\mathbf{h}}_{i,n} \triangleq [\widehat{h}_{i,n,1}, \widehat{h}_{i,n,2}, ..., \widehat{h}_{i,n,N_t}]\}_{i,n=1, i \neq n}^{N_a} \in \{\widehat{\mathbf{H}}_{i,n,:}\}_{i,n=1, i \neq n}^{N}$ appear in the *denominator* of the OF. Thus, to reach the minimum SINR, the perturbation signal $\mathbf{s}_p$ must be crafted so that $\{\widehat{\mathbf{H}}_{i,n,:}\}_{n=1}^{N}$ are *increased*. Ideally, to prevent detection, the adversary aims to keep these channel gains less than a threshold $h_{\text{th}}$ determined by the CPU, which indicates this value based on its observations from the channels' history. If the received channel gains at the CPU are higher than this value, the CPU notices abnormality in the system and prompts TXs to resend their data. If this problem occurs several times, the CPU can detect the attack occurrence. In practice, $h_{\text{th}}$ is unknown to the adversary. Hence, during the perturbation, the adversary determines the maximum value of channel gains of vertices denoted by $h_{\text{diag,max}}$ and tries to keep the gains of the perturbed channels below this new threshold. Based on the discussion above, the corresponding optimization problem is formulated in (8), which includes two additional constraints (8d) and (8e) compared to (6). Remarkably, the constraint (8d) corresponds to keeping perturbed channel gains below the determined threshold. In addition, the constraint (8e) limits $N_a$ to the minimum values between the total number of interference channels, i.e., $N^2 - N$, and $L$. The optimization problem (8) is non-convex and difficult to solve. Thus, we propose a *heuristic* algorithm to achieve a suboptimal but reasonably good solution. Similar to the $B_c$ vertex perturbation, the main part of the $B_c$ edge perturbation is the perturbation signal design. Again, it consists of two parts: The first part is similar to that of the $B_c$ vertex perturbation, while the second part is different. Let $\mathbf{s}_{p,i}$ be the corresponding perturbation signal for randomly selected $\mathbf{h}_{i,i}$. For this part, to satisfy the maximum value for the channel gain and thus to avoid detection by the CPU, $s_{p,i}$ includes an additional part with the same phase as the original channel but with an amplitude equal to $h_{\text{diag,max}}$, i.e., $h_{\text{diag,max}} e^{j \angle \mathbf{h}_{i,i}}$. Thus, the perturbation signal is $\mathbf{s}_{p,i} \leftarrow \widehat{\mathbf{h}}_{i,i} + h_{\text{diag,min}} e^{j \angle \mathbf{h}_{i,i}}$, where, $\widehat{\mathbf{h}}_{i,i}$ has been defined in the perturbation signal design of the $B_c$ vertex perturbation. The $B_c$ edge perturbation is summarized in **Algorithm 2**.

The $B_c$ perturbations assume that the adversary may perturb a number of channels, and in the worst case, all channels. These perturbations do not consider any restrictions on the adversary's power budget. However, in practice, the power budget of an adversary is limited, i.e., L2. Therefore, in the

$$\min_{\widehat{\mathbf{H}}} \quad \sum_{n \in \mathcal{A}} \mathbf{Z}_{n,N_t+1} \log_2(1 + \widehat{\text{SINR}}_n) = \sum_{n \in \mathcal{A}} \mathbf{Z}_{n,N_t+1} \log_2(1 + \frac{|\mathbf{H}_{n,n,:}^\dagger \mathbf{q}_n|^2}{\sum_{\substack{i=1 \\ i \neq n}}^{N} |\widehat{\mathbf{H}}_{i,n,:}^\dagger \mathbf{q}_i|^2 + \mathbf{Z}_{n,N_t+2}}) \quad (8a)$$

$$\text{s.t.} \quad C_0: \quad \mathcal{A} \subseteq \mathcal{N} \quad (8b)$$

$$C_1: \quad \|\mathbf{q}_i\|_2^2 \leq P_{\max}, \ \forall \ i \in \mathcal{N} \quad (8c)$$

$$C_2: \quad |\widehat{\mathbf{h}}_{i,n}| \leq h_{\text{diag,max}} \mathbf{1}_{N_t}, \ \forall \ \{\widehat{\mathbf{h}}_{i,n} \triangleq [\widehat{h}_{i,n,1}, \widehat{h}_{i,n,2}, ..., \widehat{h}_{i,n,N_t}]\}_{\substack{i=1 \\ i \neq n}}^{N_a} \in \{\widehat{\mathbf{H}}_{i,n,:}\}_{\substack{i \neq n \\ \{i,n\} \in \mathcal{N}}} \quad (8d)$$

$$C_3: \quad 1 \leq N_a \leq \min(L, N^2 - N) \quad (8e)$$

---

**Algorithm 2** $B_c$ edge perturbation

**Input:** $\mathbf{H}$, $l_c$
1: Determine $\mathcal{H}_{\text{diag}} := \{\mathbf{H}_{n,n,:}\}_{n=1}^N$, $h_{\text{diag,max}} = \max \mathcal{H}_{\text{diag}}$, and $\overline{\mathcal{H}}_{\text{diag}} := \{\mathbf{H}_{i,n,:}\}_{i \neq n, \{i,n\} \in \mathcal{N}}$,
2: $N_a = \lfloor N \times l_c \rfloor$,
3: **for** $i = 1, 2, ..., N_a$ **do**
4: $\quad$ Randomly select one element of $\overline{\mathcal{H}}_{\text{diag}}$ as $\mathbf{h}_{i,i} = |\mathbf{h}_{i,i}|e^{j\angle \mathbf{h}_{i,i}}$,
5: $\quad \widehat{\mathbf{h}}_{i,i} \leftarrow |\mathbf{h}_{i,i}|e^{j(\angle \mathbf{h}_{i,i} + \Pi)}$
6: $\quad \mathbf{s}_{p,i} \leftarrow \widehat{\mathbf{h}}_{i,i} + h_{\text{diag,max}} e^{j\angle \mathbf{h}_{i,i}}$
7: $\quad \mathbf{h}_{i,i} \leftarrow \mathbf{h}_{i,i} + \mathbf{s}_{p,i}$
8: $\widehat{\mathbf{H}} \leftarrow$ Updated $\mathbf{H}$ with the modified $\overline{\mathcal{H}}_{\text{diag}}$

**Output:** $\widehat{\mathbf{H}}$

---

following, two $B_p$ perturbations are introduced that take this constraint into consideration.

### B. $B_p$ perturbations

This subsection introduces $B_p$ vertex and $B_p$ edge perturbations. In these perturbations, both the limited power budget, i.e., L2, and the minimization of attack detection by the CPU must be satisfied. Regarding L2, we introduce a power perturbation percentage $0 \leq l_p \leq 1$, which determines the adversary's power budget, $P_a$. $l_p$ relates $P_a$ to the summation of transceiver pairs' channel gains. For instance, $l_p = 0.1$ indicates that $P_a$ is approximately 10% of the channel gains of all desired channels. The optimization problems for $B_p$ perturbations are similar to those for $B_c$ perturbations, given by (7) and (8), with an additional constraint related to L2, which is $P_a \leq \sum_{n=1}^{N} |\mathbf{H}_{n,n,:}|$.

*1) $B_p$ vertex perturbation:* Similar to the $B_c$ vertex perturbation, this perturbation reduces the SINR defined in (3) by perturbing the channel information of transceiver pairs, i.e., $\{\mathbf{H}_{n,n,:}\}_{n=1}^N$, through the design of an effective perturbation signal $\mathbf{s}_p$. To maximize the impact of perturbations on the QoC, given a limited power budget, the adversary must carefully selects channels to perturb. To do so, it perturbs the channels with the highest gain values first, leading to the highest reduction in corresponding SINRs [26]. At the beginning of its attack, the adversary checks whether the difference between the maximum and minimum channel gains, i.e., $h_{\text{diag,max}} - h_{\text{diag,min}}$, exceeds its power budget. If so, it spends its entire power budget to reduce $h_{\text{diag,max}}$. Otherwise, as long as the power budget allows, the adversary selects the channel with the highest gain, calling it $h_{\text{diag,max}}$, and perturbs it. After each perturbation, the adversary subtracts $h_{\text{diag,max}} - h_{\text{diag,min}}$ from the remaining power budget. If its power budget allows, the adversary will perturb all channel information of transceiver pairs. The perturbation signal $s_p$ used in this perturbation is similar to that of the $B_c$ vertex perturbation. **Algorithm 3** summarizes this perturbation.

---

**Algorithm 3** $B_p$ vertex perturbation

**Input:** $\mathbf{H}$, $l_p$
1: Determine $\mathcal{H}_{\text{diag}} := \{\mathbf{H}_{n,n,:}\}_{n=1}^N$, $h_{\text{diag,min}} = \min \mathcal{H}_{\text{diag}}$, and $h_{\text{diag,max}} = \max \mathcal{H}_{\text{diag}}$,
2: $P_a = l_p \times \sum_{n=1}^{N} |\mathbf{H}_{n,n,:}|$
3: **if** $h_{\text{diag,max}} - h_{\text{diag,min}} > P_a$ **then**
4: $\quad h_{\text{diag,max}} \leftarrow h_{\text{diag,max}} - P_a$
5: **else**
6: $\quad N_c \leftarrow N$
7: $\quad$ **while** $P_a \geq 0$ and $N_c \geq 0$ **do**
8: $\quad\quad i^* = \arg\max_i \mathcal{H}_{\text{diag}}$, $h_{\text{diag,max}} = \max \mathcal{H}_{\text{diag}}$
9: $\quad\quad$ **if** $P_a - (h_{\text{diag,max}} - h_{\text{diag,min}}) \geq 0$ **then**
10: $\quad\quad\quad |\mathbf{h}| \leftarrow |\mathcal{H}_{\text{diag}}[i^*]|$, $\boldsymbol{\theta} \leftarrow \angle \mathcal{H}_{\text{diag}}[i^*]$
11: $\quad\quad\quad \mathbf{s}_{p,i^*} \leftarrow |\mathbf{h}|e^{j(\boldsymbol{\theta}+\Pi)} + h_{\text{diag,min}} e^{j\boldsymbol{\theta}}$
12: $\quad\quad\quad \mathcal{H}_{\text{diag}}[i^*] \leftarrow \mathcal{H}_{\text{diag}}[i^*] + \mathbf{s}_{p,i^*}$
13: $\quad\quad\quad P_a \leftarrow P_a - (h_{\text{diag,max}} - h_{\text{diag,min}})$
14: $\quad\quad\quad N_c \leftarrow N_c - 1$
15: $\quad\quad$ **else**
16: $\quad\quad\quad$ **break**
17: $\widehat{\mathbf{H}} \leftarrow$ Updated $\mathbf{H}$ with the modified $\mathcal{H}_{\text{diag}}$

**Output:** $\widehat{\mathbf{H}}$

---

*2) $B_p$ edge perturbation:* Similar to the $B_c$ edge perturbation, this perturbation reduces the SINR defined in (3) via perturbing the interference channels, i.e., $\{\mathbf{H}_{i,n,:}\}_{i \neq n, \{i,n\} \in \mathcal{N}}$. To maximize the impact of perturbations on the QoC, given a limited power budget, the adversary must carefully selects channels to perturb. In this perturbation, it perturbs interference channels with the lowest gains first, leading to the highest reduction in the corresponding SINRs [26]. At the beginning of its attack, the adversary checks whether the difference between the maximum channel gain of transceiver pairs and the minimum channel gain of the interference channels, i.e., $h_{\text{diag,max}} - \overline{h}_{\text{diag,min}}$, is greater than its power budget. If so, it spends its entire power budget to increase $\overline{h}_{\text{diag,min}}$. Otherwise, as long as the power budget allows, the adversary selects the channel with the lowest gain, calling it $\overline{h}_{\text{diag,min}}$, and perturbs it. After each perturbation, the adversary subtracts

$h_{\text{diag,max}} - \overline{h}_{\text{diag,min}}$ from the remaining power budget. If its power budget allows, the adversary will perturb all channel information of interference channels. The perturbation signal $s_p$ is similar to that of the $B_c$ edge perturbation. **Algorithm 4** summarizes this perturbation.

---

**Algorithm 4** $B_p$ edge perturbation

**Input:** $\mathbf{H}$, $l_p$
1: Determine $\mathcal{H}_{\text{diag}} := \{\mathbf{H}_{n,n,:}\}_{n=1}^N$, $h_{\text{diag,min}} = \min \mathcal{H}_{\text{diag}}$, and $h_{\text{diag,max}} = \max \mathcal{H}_{\text{diag}}$,
2: Determine $\overline{\mathcal{H}}_{\text{diag}} := \{\mathbf{H}_{i,n,:}\}_{i \neq n, \{i,n\} \in \mathcal{N}}$, $\overline{h}_{\text{diag,min}} = \min \overline{\mathcal{H}}_{\text{diag}}$, and $\overline{h}_{\text{diag,max}} = \max \overline{\mathcal{H}}_{\text{diag}}$,
3: $P_a = l_p \times \sum_{n=1}^N |\mathbf{H}_{n,n,:}|$
4: **if** $h_{\text{diag,max}} - \overline{h}_{\text{diag,min}} > P_a$ **then**
5: $\quad \overline{h}_{\text{diag,min}} \leftarrow \overline{h}_{\text{diag,min}} + P_a$
6: **else**
7: $\quad N_c \leftarrow N$
8: $\quad$ **while** $P_a \geq 0$ and $\lfloor N_c \rfloor \geq 0$ **do**
9: $\quad\quad i^* = \arg\min_i \overline{\mathcal{H}}_{\text{diag}}$, $\overline{h}_{\text{diag,min}} = \min \overline{\mathcal{H}}_{\text{diag}}$
10: $\quad\quad$ **if** $P_a - (h_{\text{diag,max}} - \overline{h}_{\text{diag,min}}) \geq 0$ **then**
11: $\quad\quad\quad \mathcal{H}_{\text{diag}}[i^*] \leftarrow \mathcal{H}_{\text{diag}}[i^*] + \mathbf{s}_{p,i^*}$
12: $\quad\quad\quad |\mathbf{h}| \leftarrow |\overline{\mathcal{H}}_{\text{diag}}[i^*]|$, $\boldsymbol{\theta} \leftarrow \angle \overline{\mathcal{H}}_{\text{diag}}[i^*]$
13: $\quad\quad\quad \mathbf{s}_{p,i^*} \leftarrow |\mathbf{h}|e^{j(\boldsymbol{\theta}+\Pi)} + h_{\text{diag,max}}e^{j\boldsymbol{\theta}}$
14: $\quad\quad\quad \overline{\mathcal{H}}_{\text{diag}}[i^*] \leftarrow \overline{\mathcal{H}}_{\text{diag}}[i^*] + \mathbf{s}_{p,i^*}$
15: $\quad\quad\quad P_a \leftarrow P_a - (h_{\text{diag,max}} - \overline{h}_{\text{diag,min}})$
16: $\quad\quad\quad N_c \leftarrow N_c - 1$
17: $\quad\quad$ **else**
18: $\quad\quad\quad$ **break**
19: $\widehat{\mathbf{H}} \leftarrow$ Updated $\mathbf{H}$ with the modified $\overline{\mathcal{H}}_{\text{diag}}$
**Output:** $\widehat{\mathbf{H}}$

---

The performance of the four proposed adversarial attacks will be evaluated in Section VI. In addition, the following section describes how the CPU can detect these perturbations.

## V. Distribution of Channel Tensor Eigen-Values

Decomposing matrix eigenvalues provides valuable insights into its characteristics, as evidenced by prior work [58], [59]. For instance, in the context of mmWave communications, the analysis of channel eigenvalue distributions can effectively discern between perfectly estimated channels and those with imperfections [60]. Consequently, this approach proves valuable as a potential tool for detecting adversarial activity in the channel tensor. The distribution of eigenvalues is determined by the singular value decomposition (SVD). The SVD of $\mathbf{H}$ is defined as $\mathbf{H}_{\text{SVD}} = \mathbf{U}_H \Sigma_H \mathbf{V}_H^\dagger$, where $\mathbf{U}_H$ and $\mathbf{V}_H$ are unitary matrices that correspond to the left and right eigenvectors of $\mathbf{H}$, respectively. In addition, $\Sigma_H$ is a diagonal matrix with diagonal elements $\sigma_1 \geq \sigma_2 \geq \cdots \geq \sigma_{|\mathcal{V}|} \geq 0$, which are the eigenvalues of $\mathbf{H}$.

Given the eigenvalues of $\mathbf{H}$, this paper proposes the use of *One-Sample Kolmogorov-Smirnov (KS) Test* to determine their distribution. The One-Sample KS Test compares the observed Cumulative Distribution Function (CDF) for a variable with a specified theoretical distribution, such as normal, uniform, Poisson, etc. This test is only valid for continuous distributions. It computes *KS statistic*, which is the largest difference (in absolute value) between the observed and theoretical CDFs. The KS statistic for a given CDF $Q(\sigma_i)$ is defined as:

$$D = \max_{\sigma_i \in \mathcal{X}} |P_\mathcal{X}(\sigma_i) - Q(\sigma_i)|, \quad (9)$$

where, $\mathcal{X} = \{\sigma_i\}_{i=1}^{|\mathcal{V}|}$ is an ordered observations with $\sigma_{|\mathcal{V}|} \leq \sigma_2 \leq ... \leq \sigma_1$. The empirical CDF of these samples, $P_\mathcal{X}(x)$, is determined as:

$$P_\mathcal{X}(x) = \begin{cases} 0, & x < \sigma_{|\mathcal{V}|}, \\ \frac{k}{S}, & \sigma_k \leq x \leq \sigma_{k+1}, \\ 1, & x \geq \sigma_1, \end{cases} \quad (10)$$

The One-Sample KS Test checks whether $\mathcal{X}$ could reasonably come from the hypothesized distribution $Q(\sigma_i)$ [61]. It has been proven that if the observations (in this case, eigenvalues) come from the distribution $Q(\sigma_i)$, then $D$ converges to 0 almost surely in the limit as the number of samples approaches to infinity [62]–[64]. However, an infinite number of samples is not available in practice. Instead, if $D$ is less a predefined critical value $D_{\text{crit}}$, it indicates that the sample data follows $Q(\sigma_i)$. The critical value for the known population mean and standard deviation is determined through One-Sample KS table [65]. However, in our scenario, the mean of the population and the standard deviation of the samples are unknown and need to be estimated from the samples. In this case, $D_{\text{crit}}$ is determined by the Lilliefors Test table [66], [67]. Given $D_{\text{crit}}$, the best fitting distribution of eigenvalues corresponds to the minimum value of the KS statistic $D$ among all different hypothesized distributions. Thus, we define the optimization problem as follows:

$$Q^* = \arg\min_{Q \in \mathcal{Q}} \max_{\sigma_i \in \mathcal{X}} |P_\mathcal{X}(\sigma_i) - Q(\sigma_i)|,$$
$$\text{s.t.} \max_{\sigma_i \in \mathcal{X}} |P_\mathcal{X}(\sigma_i) - Q(\sigma_i)| \leq D_{\text{crit}}, \quad (11)$$

where, $\mathcal{Q}$ is the set of CDFs that includes all hypothesized continuous distributions. In addition, the distribution $Q^*$ corresponds to the minimum value of $D$. The corresponding results are presented in Section VI. They show that the eigenvalues of the channel tensor $\mathbf{H}$ follow the Johnson's SU distribution, as the minimum value of $D$ belongs to this distribution. The Johnson's SU distribution is a continuous probability distribution with four real-valued parameters, and its probability density function (PDF) is defined as:

$$p_J(x) = \frac{\delta}{\lambda\sqrt{2\pi}} \frac{1}{\sqrt{1+(\frac{x-\xi}{\lambda})^2}} e^{-\frac{1}{2}(\gamma+\delta\sinh^{-1}(\frac{x-\xi}{\lambda}))^2}, \quad (12)$$

where, $\lambda > 0$ and $\delta > 0$. In addition, $\gamma$ and $\xi$ are known as the shape parameters [37].

## VI. Performance Analysis

This section evaluates the performance of the proposed adversarial attacks on the total QoC of the system, as well as on the distribution of $\mathbf{H}$'s eigenvalues.

We generated our own input datasets by simulating transceiver pairs within a defined rectangular area. The transmitters are randomly placed in this area, and for each transmitter, a corresponding receiver is uniformly distributed within a set distance range, specifically $[d_{\min}, d_{\max}]$. The channels utilized in our simulation were generated based on the channel model provided in [68], where the channel between TX $i$ and RX $j$ is represented as $\mathbf{h}_{j,i} = 10^{-L(d_{ji})/20}\sqrt{\psi_{ji}\rho_{ji}}\mathbf{g}_{ji}$, $\forall\ i,j \in \mathcal{N}$. In this model, $L(d_{ji}) = 148.1 + 37.6 \log_2(d_{ji})$ corresponds to the path-loss at a distance of $d_{ji}$ in kilometers. Additionally, $\psi_{ji} = 9$ dBi, $\rho_{ji}$ represents the antenna gain and shadowing coefficient, respectively, while $\mathbf{g}_{ji}$ follows a small-scale fading coefficient that conforms to $\mathcal{CN}(\mathbf{0}_{N_t}, \mathbf{I}_{N_t})$.

In order to minimize the CSI training overhead, we made an assumption that channels are only available for those transceiver pairs where the distance between each transmitter and its corresponding receiver is less than 500 meters. For our data split ratio, we used 2000 samples for training and 500 samples for testing. Each sample included $N$ transceiver pairs, which were used to evaluate the proposed adversarial attacks. We employ the same GNN architecture as in [15] and [69], which is a 3-layer graph neural network. As outlined in Section III-A, the network takes the channel states $\{\mathbf{Z}_{n,1:N_t}^{\dagger}\}_{n=1}^{N}$ and users' weights $\{w_n\}_{n=1}^{N}$ as input, and produces the beamforming vectors of users as output. The loss function at the last layer of the GNN is defined in (4). We use Adam [70] as the GNN optimizer, with a learning rate of 0.001. The number of transceiver pairs $N$ and SNR are kept the same for both the training and testing phases.

First, we analyze the output of optimization problem (11) to determine the distribution of $\mathbf{H}$'s eigenvalues before applying adversarial attacks. We set up 25 wireless networks with N = 40 pairs of transceivers distributed randomly in an area of 1000 square meters. Thus, each generation of wireless communication results in 40 eigenvalues and thus, the sample size is $25 \times 40 = 1000$. In addition, each RX is uniformly distributed in $[d_{\min}, d_{\max}] = [10, 50]$ from its corresponding TX. As a part of (11), we need to determine $D_{\text{crit}}$, which depends on the number of eigenvalues and $\alpha$ in the Lilliefors Test Table. $\alpha$ is associated with the probability of $D \leq D_{\text{crit}}$, where $P(D \leq D_{\text{crit}}) = 1 - \alpha$. As $\alpha$ increases, $D_{\text{crit}}$ decreases. The minimum and maximum values of $\alpha$ in the Lilliefors Test Table are 0.01 and 0.2, respectively. In this study, we consider $\alpha = 0.01$ to determine $D_{\text{crit}}$. According to this $\alpha$, $D_{\text{crit}}$ is determined as:

$$D_{\text{crit}} = \frac{1.035}{\frac{N_\sigma + 0.83}{\sqrt{N_\sigma}} - 0.01}, \tag{13}$$

where, $N_\sigma$ denotes the sample size used to determine the distribution of the eigenvalues. In our case, $N_\sigma = 1000$, resulting in $D_{\text{crit}} = 0.0327$. Table I compares the KS statistic of 84 continuous distributions. Abbreviation for the distributions used in this table are listed in Appendix A. This table shows that the minimum value of the KS statistic corresponds to Johnson's SU distribution. In addition, this minimum value satisfies the constraint in (11), since $0.0296 < D_{\text{crit}} = 0.0327$. Thus, we conclude that the eigenvalues of $\mathbf{H}$ follow this distribution.

Next, Table II evaluates (11) for different values of $N$ and $N_\sigma$, where Johnson's SU distribution achieved the best outcomes in most cases. For distributions with better KS Statistic values, the best fitted distributions are listed in order, from the best fit distribution to Johnson's SU distribution, along with the corresponding values. In these cases, the difference between the KS statistic values for the best fit distribution and Johnson's SU distribution is very small and insignificant.

For instance, for $N = 50$, the Johnson's SU's distribution is the second most appropriate distribution, with a KS statistic value difference of less than 0.003 compared to Non-central Student's t distribution. Moreover, Johnson's SU distribution is always the best distribution of eigenvalues at high $N_\sigma$ values across all $N$ values.

The results of Tables I and II lead to the conclusion that the eigenvalues of $\mathbf{H}$ follow Johnson's SU distribution.

### A. Impact of Adversarial Attacks

In this section, we first evaluate the performance of GNN in terms of QoC under the absence of any attacks. Next, we analyze the effect of the proposed adversarial attacks presented in **Algorithms 1**, **2**, **3**, and **4** on both the total QoC of the system and the distribution of $\mathbf{H}$'s eigenvalues.

*1) QoC of the system with no attack:* To evaluate the GNN performance, different scenarios with a varying number of vertices, ranging from 20 to 50, and three different states for the distance between transmitters (TXs) and receivers (RXs) are considered in Table III. This table reveals that GNN performance improves as both the number of users and the distance between TX and RX increase.

*2) QoC of the perturbed system:* Many different scenarios are considered to assess the effect of adversarial attacks on the total QoC, i.e., $\sum_{n=1}^{N} \mathbf{Z}_{n,N_t+1}\log_2(1 + \text{SINR}_n)$.

It's important to note that when the adversary uses either the $B_c$ or $B_p$ vertex perturbation, it affects $N$ channels corresponding to the desired channels. When it uses $B_c$ and $B_p$ edge perturbations, however, it affects $N^2 - N$ channels, which correspond to interference channels. Therefore, $B_c$ and $B_p$ edge perturbations affect $N^2 - 2N$ more channels compared to $B_c$ and $B_p$ vertex perturbations, leading to a more severe decrease in the total QoC. To make a fair comparison, when the adversary uses $B_c$ or $B_p$ edge perturbation, it considers $l_c$ or $l_p$ divided by $N - 1$, rather than $l_c$ or $l_p$.

We introduce three heuristic perturbations for comparison with the proposed adversarial perturbations. They are as follows: I) *Upper Bound $B_c$ Perturbation*: In this attack, the adversary prioritizes L1 and targets the selected channel with the highest power, effectively setting its value to zero. This attack serves as an upper benchmark for $B_c$ vertex perturbation. II) *Single $B_c$ Perturbation*: The adversary focuses on attacking a single randomly selected channel, abiding by constraint L1 with $|L1| = 1$. This attack zeroes the chosen channel's information, regardless of the energy required. III) *Uniform $B_p$ Perturbation*: This perturbation follows constraint L2 and perturbs the channel information of transceiver pairs. The adversary selects $l_p$ channels and transmits a perturbation signal to decrease the magnitude of these channels by the same amount, determined by the available power budget over the number of selected channels. Note that no design of the perturbation signal $\mathbf{s}_p$ is involved in these proposed heuristic perturbations. The paper includes a table that displays the QoC of the system after applying these perturbations, with lower values indicating better perturbation effectiveness.

TABLE I: KS Statistic Values for Different Distributions with No Adversarial Attacks

| | | | | | | | |
|---|---|---|---|---|---|---|---|
| Alpha | 1.0 | Folded Cauchy | 0.5137 | Hyperbolic Secant | 0.0544 | Normal | 0.2782 |
| Anglit | 1.0 | Folded Normal | 0.3257 | Inverted Gamma | 1.0 | Pareto | 0.1189 |
| Arcsine | 0.7854 | Weibull$_{min}$ | 0.1058 | Inverse Normal | 0.0489 | Pearson type III | 0.1024 |
| Beta | 0.1028 | Weibull$_{max}$ | 0.7783 | Inverted Weibull | 0.0689 | Power Law | 0.4085 |
| Beta Prime | 0.9747 | Generalized Logistic | 0.1592 | Johnson SB | 0.0521 | Power Log Normal | 0.0482 |
| Bradford | 0.6453 | Generalized Pareto | 0.1356 | Johnson's SU | **0.0296** | Power Normal | 1.0 |
| Burr | 0.8095 | Generalized Norm | 0.2545 | KStwobign | 0.2269 | R-distributed | 1.0 |
| Cauchy | 0.2067 | Generalized Exponential | 0.1 | Laplace | 0.2482 | Reciprocal | 1.0 |
| Chi | 0.1786 | Generalized Extreme Value | 0.0679 | Lévy | 1.0 | Rayleigh | 0.2517 |
| Chi-squared | 0.1025 | Gauss Hypergeometric | 0.2892 | Left-skewed Lévy | 0.666 | Rice | 0.2517 |
| Cosine | 1.0 | Gamma | 0.6183 | Logistic | 0.1913 | Recipinvgauss | 0.0885 |
| Double Gamma | 0.3161 | Generalized Gamma | 0.0612 | Log Gamma | 0.3153 | Semicircular | 1.0 |
| Double Weibull | 0.2635 | Generalized Half-Logistic | 0.1692 | Log Laplace | 0.7857 | Singh-Maddala | 0.6566 |
| Erlang | 0.552 | Gilbrat | 0.1668 | Log Normal | 0.0535 | Student's t | 0.1895 |
| Exponential | 0.1 | Gompertz | 0.9755 | Lomax | 0.1238 | Triangular | 0.6921 |
| Exponentiated Normal | 0.127 | Gumbel$_r$ | 0.3073 | Maxwell | 0.2301 | Truncated Exponential | 0.6982 |
| Exponentiated Weibull | 0.0954 | Gumbel$_l$ | 0.5728 | Mielke's Beta-Kappa | 0.5746 | Truncated Normal | 1.0 |
| Exponentiated Power | 0.2106 | Half Cauchy | 0.1878 | Nakagami | 0.1699 | Tukey-Lambda | 0.2082 |
| F | 0.6013 | Half Logistic | 0.3607 | Non-central Chi-squared | 0.1181 | Uniform | 0.89 |
| Fatigue Life | 0.0692 | Half Normal | 0.7887 | Non-Central F | 0.1567 | Von Mises Line | 1.0 |
| Fisk | 0.7956 | HalfGenNorm | 0.2152 | Non-central Student's t | 0.0423 | Wald | 0.0714 |

TABLE II: KS Statistic Values for Different Number of Users and Eigenvalues with No Adversarial Attacks

| | $N_\sigma$ | | | | | | | |
|---|---|---|---|---|---|---|---|---|
| | 100 | 200 | 300 | 500 | 1000 | 2000 | 3000 | 5000 |
| $N$ \ $D_{\text{crit}}$ | 0.102 | 0.073 | 0.06 | 0.046 | 0.033 | 0.023 | 0.019 | 0.015 |
| 10 | 0.056 | 0.054 | 0.057 | 0.057 | 0.039 | 0.038 | 0.038 | 0.038 |
| 20 | invnor[1]: 0.038<br>genext[2]: 0.05<br>falife[3]: 0.050<br>JSU[4]: 0.052 | 0.059 | 0.046 | 0.041 | 0.036 | 0.032 | 0.032 | 0.031 |
| 30 | 0.043 | 0.045 | 0.059 | 0.048 | 0.038 | 0.033 | 0.032 | 0.033 |
| 40 | hcauchy[5]: 0.081<br>JSU: 0.110 | Gilbrat: 0.044<br>inv-nor: 0.053<br>exweib[6]: 0.054<br>JSU: 0.055 | plognorm[7]: 0.054<br>JSU: 0.056 | invnor: 0.067<br>JSU: 0.071 | 0.030 | 0.030 | 0.031 | 0.028 |
| 50 | 0.035 | nct[8]: 0.043<br>JSU: 0.045 | nct: 0.053<br>JSU 0.056 | nct: 0.054<br>JSU: 0.055 | 0.045 | 0.029 | 0.029 | 0.030 |

[1]Inverse Normal, [2]Generalized Extreme Value, [3]Fatigue Life, [4]Johnson's SU, [5]Half Cauchy, [6]Exponentiated Weibull, [7]Power Log Normal, [8]Non-central Student's t.

TABLE III: The total QoC with no attack

| | [$d_{\min}, d_{\max}$] | | |
|---|---|---|---|
| N | [30, 30] | [10, 50] | [2, 65] |
| 20 | 121.80 | 122.93 | 127.44 |
| 30 | 144.38 | 157.56 | 157.50 |
| 40 | 164.56 | 177.66 | 189.63 |
| 50 | 179.52 | 197.29 | 214.41 |

First, Table IV evaluates the total QoC for different values of $N$, $l_c$, and $l_p$, where SNR=10 dB, and the table entries represent the total QoC after applying various attacks normalized by the total QoC before implementing attacks. There are several points to note in this table: 1) As expected, the performance of adversarial attacks increases with increasing $l_c$ or $l_p$, 2) The best and worst performances belong to $B_c$ edge and $B_p$ edge perturbations, respectively. The $B_c$ edge perturbation leads to a reduction of approximately 95% of the total QoC, which corresponds to $N = 30$ and $l_p = 0.9$. While, in the best case, $B_p$ edge perturbation can achieve only 6% reduction in the total QoC, which is for $N = 50$ and $l_p = 0.9$, and 3) In most cases, attacks are more effective as $N$ increases. However, the $B_c$ edge perturbation shows a different pattern, and its performance mainly declines as $N$ increases.

Furthermore, the following are the results of the proposed heuristic perturbations; The Upper Bound $B_c$ perturbation consistently outperforms the $B_c$ vertex, $B_p$ vertex, and $B_p$ edge perturbations across various network sizes and perturbation levels. However, it exhibits lower performance compared to the $B_c$ edge perturbation in most scenarios. This highlights the importance of constraint (7d) on the effectiveness of the proposed perturbations and the risk of detection after the initial

TABLE IV: The total QoC after Applying Adversarial Attacks for Different $N$

| | $N = 20$ | | | | | $N = 30$ | | | | | $N = 40$ | | | | | $N = 50$ | | | | |
|---|---|---|---|---|---|---|---|---|---|---|---|---|---|---|---|---|---|---|---|---|
| $l_c$ or $l_p \Longrightarrow$ | 0.1 | 0.3 | 0.5 | 0.7 | 0.9 | 0.1 | 0.3 | 0.5 | 0.7 | 0.9 | 0.1 | 0.3 | 0.5 | 0.7 | 0.9 | 0.1 | 0.3 | 0.5 | 0.7 | 0.9 |
| $B_c$ vertex | 0.966 | 0.932 | 0.879 | 0.838 | 0.782 | 0.941 | 0.897 | 0.845 | 0.794 | 0.739 | 0.911 | 0.862 | 0.818 | 0.754 | 0.703 | 0.879 | 0.847 | 0.781 | 0.735 | 0.66 |
| $B_c$ edge | 0.827 | 0.571 | 0.429 | 0.305 | 0.202 | 0.597 | 0.236 | 0.129 | 0.082 | 0.056 | 0.717 | 0.436 | 0.246 | 0.137 | 0.079 | 0.805 | 0.583 | 0.424 | 0.31 | 0.214 |
| $B_p$ vertex | 0.99 | 0.963 | 0.900 | 0.821 | 0.775 | 0.975 | 0.933 | 0.870 | 0.779 | 0.719 | 0.956 | 0.918 | 0.834 | 0.738 | 0.675 | 0.957 | 0.894 | 0.814 | 0.695 | 0.649 |
| $B_p$ edge | 0.961 | 0.949 | 0.955 | 0.949 | 0.944 | 0.967 | 0.959 | 0.948 | 0.959 | 0.957 | 0.964 | 0.960 | 0.960 | 0.946 | 0.944 | 0.955 | 0.949 | 0.952 | 0.948 | 0.942 |
| Upper Bound $B_c$ | 0.902 | 0.712 | 0.522 | 0.295 | 0.119 | 0.897 | 0.686 | 0.450 | 0.222 | 0.103 | 0.884 | 0.683 | 0.497 | 0.288 | 0.049 | 0.868 | 0.675 | 0.480 | 0.286 | 0.100 |
| Single $B_c$ | | | 0.999 | | | | | 0.992 | | | | | 0.978 | | | | | 0.974 | | |
| Uniform $B_p$ | 0.939 | 0.956 | 0.949 | 0.947 | 0.947 | 0.965 | 0.958 | 0.966 | 0.962 | 0.972 | 0.956 | 0.946 | 0.962 | 0.950 | 0.965 | 0.951 | 0.966 | 0.964 | 0.949 | 0.949 |

TABLE V: The total QoC after Applying Adversarial Attacks for Different TX-RX distance

$N = 20$

| | $d_{\min} = 30, d_{\max} = 30$ | | | | | $d_{\min} = 10, d_{\max} = 50$ | | | | | $d_{\min} = 2, d_{\max} = 65$ | | | | |
|---|---|---|---|---|---|---|---|---|---|---|---|---|---|---|---|
| $l_c$ or $l_p \Longrightarrow$ | 0.1 | 0.3 | 0.5 | 0.7 | 0.9 | 0.1 | 0.3 | 0.5 | 0.7 | 0.9 | 0.1 | 0.3 | 0.5 | 0.7 | 0.9 |
| $B_c$ vertex | 0.989 | 0.985 | 0.986 | 0.989 | 0.985 | 0.966 | 0.932 | 0.879 | 0.838 | 0.782 | 0.943 | 0.875 | 0.821 | 0.752 | 0.679 |
| $B_c$ edge | 0.811 | 0.562 | 0.422 | 0.288 | 0.207 | 0.827 | 0.571 | 0.429 | 0.305 | 0.202 | 0.822 | 0.572 | 0.427 | 0.302 | 0.218 |
| $B_p$ vertex | 0.989 | 0.99 | 0.987 | 0.989 | 0.993 | 0.99 | 0.963 | 0.900 | 0.821 | 0.775 | 1.00 | 0.99 | 0.953 | 0.901 | 0.775 |
| $B_p$ edge | 0.949 | 0.932 | 0.931 | 0.933 | 0.931 | 0.961 | 0.949 | 0.955 | 0.949 | 0.944 | 0.952 | 0.955 | 0.952 | 0.954 | 0.949 |

TABLE VI: The total QoC after Applying Adversarial Attacks for Different SNR

$N = 20$

| | -5 | | | | | 0 | | | | | 5 | | | | | 10 | | | | |
|---|---|---|---|---|---|---|---|---|---|---|---|---|---|---|---|---|---|---|---|---|
| SNR (dB) $\Longrightarrow$ | | | | | | | | | | | | | | | | | | | | |
| $l_c$ or $l_p \Longrightarrow$ | 0.1 | 0.3 | 0.5 | 0.7 | 0.9 | 0.1 | 0.3 | 0.5 | 0.7 | 0.9 | 0.1 | 0.3 | 0.5 | 0.7 | 0.9 | 0.1 | 0.3 | 0.5 | 0.7 | 0.9 |
| $B_c$ vertex | 0.967 | 0.919 | 0.882 | 0.836 | 0.788 | 0.967 | 0.926 | 0.879 | 0.836 | 0.786 | 0.971 | 0.929 | 0.882 | 0.838 | 0.784 | 0.966 | 0.932 | 0.879 | 0.838 | 0.782 |
| $B_c$ edge | 0.830 | 0.565 | 0.430 | 0.303 | 0.206 | 0.818 | 0.561 | 0.431 | 0.298 | 0.207 | 0.827 | 0.569 | 0.434 | 0.303 | 0.205 | 0.827 | 0.571 | 0.429 | 0.305 | 0.202 |
| $B_p$ vertex | 0.993 | 0.957 | 0.903 | 0.824 | 0.769 | 0.996 | 0.955 | 0.902 | 0.822 | 0.772 | 0.996 | 0.961 | 0.899 | 0.825 | 0.771 | 0.99 | 0.963 | 0.900 | 0.821 | 0.775 |
| $B_p$ edge | 0.961 | 0.949 | 0.951 | 0.950 | 0.943 | 0.956 | 0.951 | 0.948 | 0.951 | 0.945 | 0.954 | 0.957 | 0.950 | 0.940 | 0.944 | 0.961 | 0.949 | 0.955 | 0.949 | 0.944 |

attack. The Single $B_c$ perturbation is the mildest among all perturbations, resulting in the least reduction in the QoC of the system. This is logical since it nullifies the channel information for only a single channel. The Uniform $B_p$ vertex perturbation performs well when the number of selected channels ($l_c$ or $l_p$) is low, often matching the effectiveness of the $B_p$ vertex and $B_p$ edge perturbations. However, as the number of selected channels increases, its performance deteriorates significantly, emphasizing the importance of designing the perturbation signal $\mathbf{s}_p$ and thoughtfully selecting channels to perturb.

Next, Table V evaluates the results for different TX-RX distances, i.e., $[d_{\min}, d_{\max}]$. Three systems are considered, where RXs are uniformly distributed within $[30, 30]$, $[2, 65]$, and $[10, 50]$ distances from their corresponding TXs. Here, $N = 20$ and SNR = 10 dB. The table shows that the performance of the $B_c$ vertex perturbation increases with the increase of $d_{\max} - d_{\min}$. This indicates that this perturbation further damages the total QoC when RXs are located at significantly different distances from their TXs. On the other hand, the other perturbations do not show any specific pattern. In addition, the best performance for $B_c$ and $B_p$ edge perturbations occurs in $[d_{\min}, d_{\max}] = [30, 30]$. This implies that edge perturbations achieve their best performance in homogeneous networks.

Next, Table VI evaluates the results for various SNR values, where $N = 20$ and $[d_{\min}, d_{\max}] = [10, 50]$. This table illustrates that the proposed attacks perform similarly in both high and low SNR regimes, indicating that their performance is independent of SNR.

Next, we compare the performance of the attacks where the adversary does not care about detection. Its sole purpose is to maximize the destruction of the total QoC during the first attack, knowing that it will be detected thereafter. To achieve this, the adversary transmits the perturbation signal $\mathbf{s}_p$ to force channels to zero, without any restriction on the minimum channel gain value, i.e., the constraint C2 in (7d), when using vertex perturbations, and without any restriction on the maximum channel gain value, i.e., the constraint C1 in (8c), when using edge perturbations. Here, $N = 20$, $[d_{\min}, d_{\max}] = [10, 50]$, and SNR = 10 dB. Fig. 2 illustrates that the $B_c$ vertex perturbation results in a linear decrease in the total QoC as $l_c$ increases, when there are no limits on the minimum channel gain value. Note that the right and left vertical axes correspond to the blue and black graphs, respectively. In this figure, the difference between the $B_c$ vertex perturbations - plotted in black - increases as $l_c$ increases. For example, the $B_c$ vertex perturbation with no limit on the minimum channel gain value results in a 95% reduction in the total QoC, compared to approximately 20% reduction with the imposed limit. On the other hand, the $B_c$ edge perturbations - represented by the blue graphs - exhibit nearly identical performance.

*3) Eigenvalues' distribution of the perturbed system:* In this section, we investigate the impact of adversarial perturbations on the distribution of the channel's eigenvalues.

To do so, after applying perturbations to the channel tensor $\mathbf{H}$ of the system, SVD is used to determine the corresponding eigenvalues of the perturbed version of $\mathbf{H}$, denoted as $\widehat{\mathbf{H}}$. By obtaining the eigenvalues, the same process as described in Section V is used to solve the optimization problem (11) and determine the resulting distribution of eigenvalues. For the upcoming simulations, we set $N = 50$, $[d_{\min}, d_{\max}] = [10, 50]$,

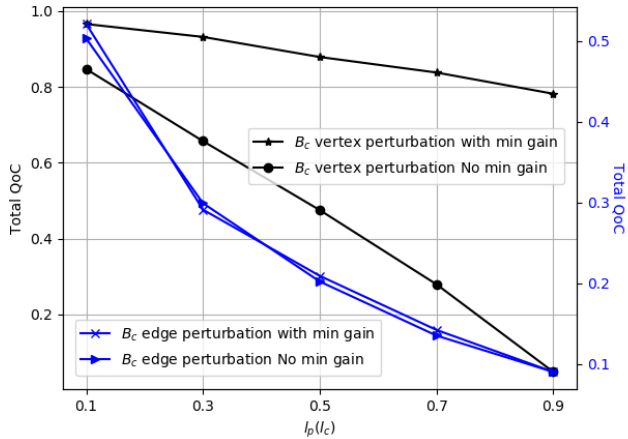

Fig. 2: Comparison $B_c$ perturbations with and without channel gain limits, $N = 20$.

and SNR = 10 dB.

First, we evaluate the impact of the $B_c$ vertex perturbation for different values of $l_c$. The output of the optimization problem (11) for the eigenvalues of $\widehat{\mathbf{H}}$ reveals that this perturbation does not alter the type of eigenvalues' distribution, but rather only modifies the distribution parameters. Fig. 3 shows how the parameters change with different $l_c$ values. This figure demonstrates that the magnitude of parameters $\lambda$ and $\delta$ decreases after applying the perturbation, compared to their values before perturbation, and they continue to decline as $l_c$ increases. On the other hand, parameters $\gamma$ and $\xi$ reach their minimum magnitude when no perturbation is present, and their magnitude increases as $l_c$ increases. Fig.4 compares the PDF of $\mathbf{H}$'s eigenvalues before and after applying the $B_c$ vertex perturbation for different values of $l_c$. This figure shows that the perturbation causes the PDF of eigenvalues to concentrate and shift upwards around a specific eigenvalue. This shift is a result of changing the corresponding parameters of the distribution, as depicted in Fig. 3. The amount of the shift increases with higher values of $l_c$. In addition, these changes lead to the CDF approaching its maximum value more rapidly and at smaller eigenvalues, as shown in Fig. 5.

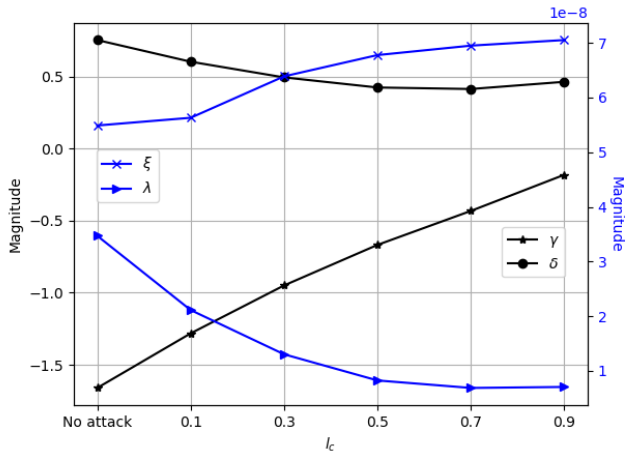

Fig. 3: Variation of Johnson's SU distribution parameter before and after applying the $B_c$ vertex perturbation.

Next, we evaluate the impact of the $B_c$ edge perturbation for different values of $l_c$. Notably, the optimization problem (11) yields distinct outcomes compared to the $B_c$ vertex perturbation. The results show that the $B_c$ edge perturbation alters the distribution of eigenvalues. For $l_c$ values of 0.1, 0.3, 0.5, 0.7, and 0.9, the eigenvalues of $\widehat{\mathbf{H}}$ follow Noncentral Student's t, Generalized Inverse Gaussian, Generalized Extreme Value, Generalized Extreme Value, and the upper half of a generalized normal distributions, respectively. The corresponding PDFs and CDFs are depicted in Fig. 6 and Fig. 7, respectively. Fig. 6 shows that the PDFs of $\widehat{\mathbf{H}}$ exhibit a peak at very small values of $\delta$. Consequently, the CDFs of $\widehat{\mathbf{H}}$ reach their maximum values more rapidly compared to the CDF of $\mathbf{H}$, as shown in Fig. 6.

Next, we assess the impact of the $B_p$ vertex perturbation. Like the $B_c$ vertex perturbation, applying this perturbation does not alter the distribution type of eigenvalues. However, it affects the parameters of the eigenvalues' distribution. Fig. 8 compares the magnitude of these parameters before and after applying this perturbation, considering various values of $l_p$.

The changes of the parameters depicted in this figure differ from the changes observed after applying the $B_c$ vertex perturbation. Specifically, the variations of $\delta$ and $\lambda$ are negligible for $l_p \leq 0.3$. In addition, these values decrease as $l_p$ increases for $l_p > 0.3$. On the other hand, the magnitude of $\xi$ decreases for $l_p \leq 0.3$, but increases as $l_p$ increases for $l_p > 0.3$. In addition, $\gamma$ exhibits a continuous increase with the growth of $l_p$. However, this increase is negligible for $l_p \leq 0.3$. Fig.9 compares the PDF of $\mathbf{H}$'s eigenvalues before and after applying the $B_p$ vertex perturbation. This figure shows that applying this perturbation causes the PDF to become more concentrated and shifts it up around different eigenvalues for different $l_p$ values. For values of $l_p$ up to 0.3, the shift is negligible, but it becomes more pronounced as $l_p$ increases. In addition, the corresponding CDF approaches its maximum value more quickly and at smaller eigenvalues compared to that with no perturbation as shown in Fig. 10.

Next, we evaluate the effect of the $B_p$ edge perturbation. Similar to the $B_c$ and $B_p$ vertex perturbations, the solution of the optimization problem (11) for eigenvalues of $\widehat{\mathbf{H}}$ shows that these eigenvalues follow Johnson's SU distribution. This perturbation alters the parameters of the eigenvalues' distribution. Fig. 11 compares the magnitude of these parameters before and after applying this perturbation for different values of $l_p$. The application of the perturbation decreases $\xi$ and increases $\gamma$ values. In addition, as $l_p$ increases, the $\xi$ value also increases, while the $\gamma$ value decreases. However, this trend does not apply to the $\delta$ and $\lambda$ values, as they remain unchanged before and after the perturbation. The PDFs of $\mathbf{H}$'s eigenvalues before and after the application of the $B_p$ edge perturbation are compared in Fig.12. In contrast to the $B_c$ and $B_p$ vertex perturbations, this figure shows that the application of the $B_p$ edge perturbation causes a *downward* shift in the distribution. The magnitude of this shift increases as $l_p$ increases. Additionally, Fig. 13 shows that there is no significant difference between the CDF of the eigenvalues before and after applying the perturbation.

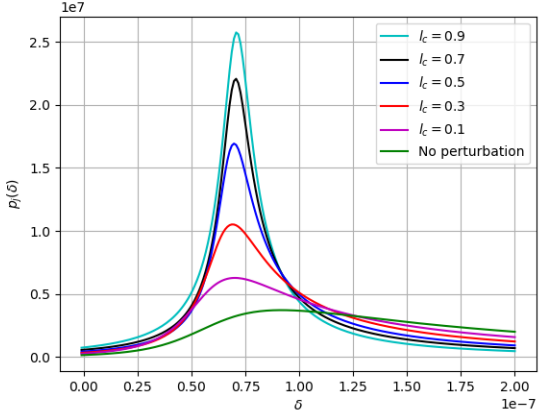

Fig. 4: PDF for EVs after $B_c$ vertex perturbation.

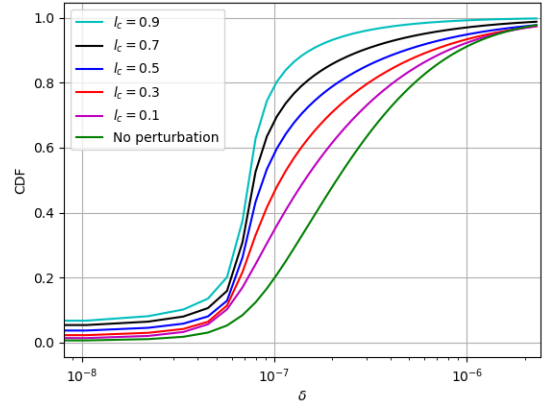

Fig. 5: CDF for EVs after $B_c$ vertex perturbation.

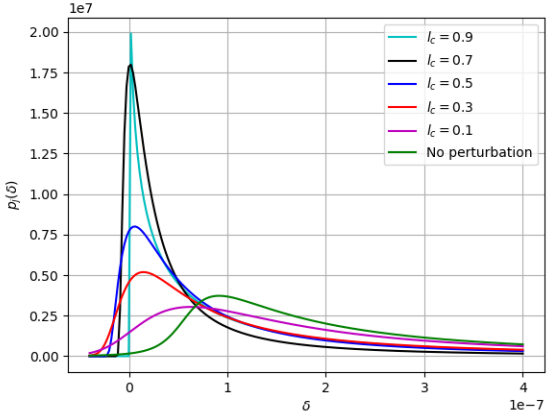

Fig. 6: PDF for EVs after $B_c$ edge perturbation.

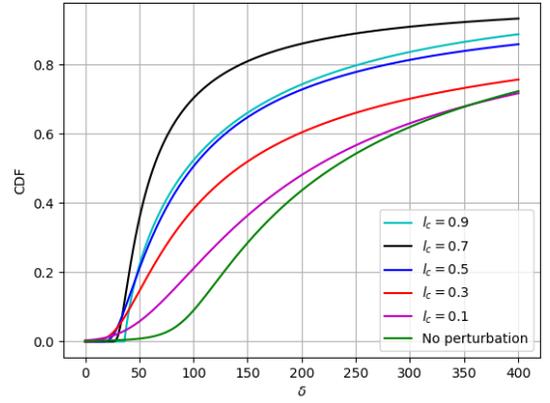

Fig. 7: CDF for EVs after $B_c$ edge perturbation.

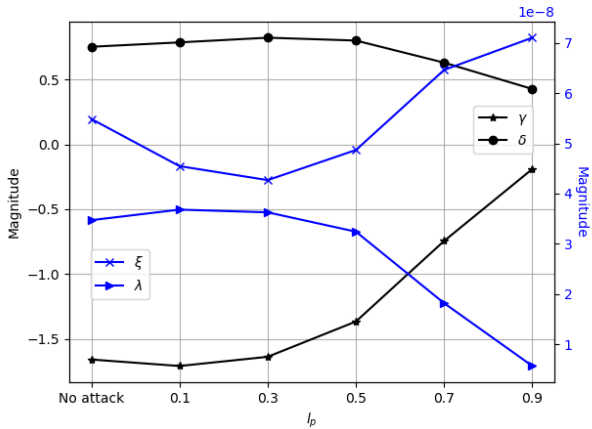

Fig. 8: Variation of Johnson's SU distribution parameter before and after applying the $B_p$ vertex perturbation.

Fig. 14 illustrates the location of the maximum value of the PDF before and after the application of the perturbations. This figure indicates the impact of adversarial perturbations on the eigenvalue at which the PDF reaches its maximum value. The maximum value is observed at $\delta = 9.15e-8$ without adversarial perturbation, and this value reduces when adversarial perturbations are applied. Notably, the reduction is more prominent after applying the $B_c$ perturbations compared to the $B_p$ perturbations. Furthermore, the amount of change in the values of $l_c$ and $l_p$ is nearly equal when the $B_c$ vertex and $B_p$ edge perturbations are applied.

Our research makes three main assumptions that might limit its real-world applicability. 1) We haven't factored in the effect of the adversary-to-CPU channel. This might limit our proposed attacks' effectiveness, especially in non-controlled environments. 2) We assumed perfect CSI knowledge by the adversary, while in reality, this information could be inaccurate, limiting the attack's success. 3) We focused on a single adversary, neglecting scenarios involving multiple adversaries.

These assumptions are more valid under certain conditions, such as controlled environments with minimal interference and change, static wireless environments, and less complex networks. However, in more dynamic real-world applications, many factors can influence the adversary's signal, the accuracy of CSI, and the potential for multiple adversaries. These factors might limit the effectiveness of an attack, indicating the need for future research to consider these variables for better real-world applicability.

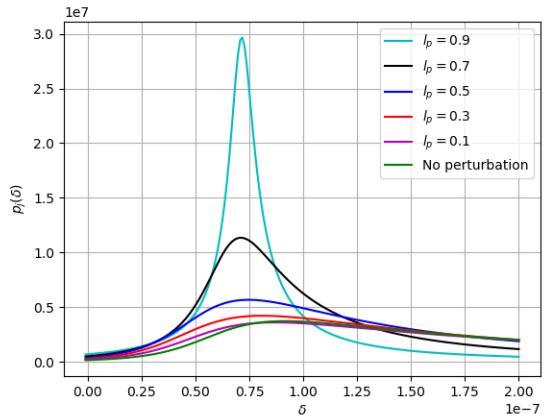

Fig. 9: PDF for EVs after $B_p$ vertex perturbation.

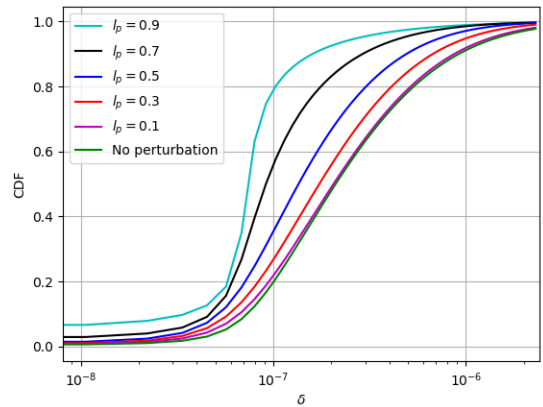

Fig. 10: CDF for EVs after $B_p$ vertex perturbation.

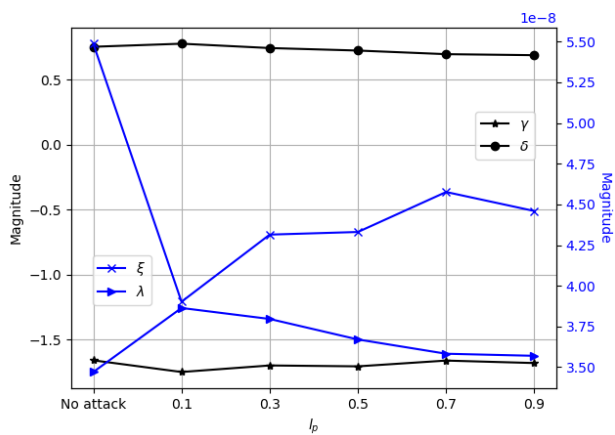

Fig. 11: Variation of Johnson's SU distribution parameter before and after applying the $B_p$ edge perturbation.

## VII. CONCLUSION

This paper considers adversarial attacks on centralized GNN-based peer-to-peer communication. Four new adversarial attacks, namely $B_c$ and $B_p$ vertex disruptions, $B_c$ and $B_p$ edge disruptions, are proposed. Additionally, the distribution of eigenvalues of the channel tensor $\mathbf{H}$ before and after adversarial attacks is determined. The performance of the proposed attacks is evaluated based on the total QoC and distribution of eigenvalues. The results of the total QoC analysis validate the effectiveness of the attacks across different numbers of users, SNR levels, and adversary power budgets. Furthermore, the analysis reveals that the $B_c$ edge perturbation exhibits the most severe impact on the total QoC, resulting in a 95% reduction, while the $B_p$ edge perturbation shows the least impact on the total QoC, leads to a 6% reduction. Moreover, except for the $B_c$ edge perturbation, the effectiveness of all attacks increases with the number of transceiver pairs ($N$). The $B_c$ vertex perturbation significantly impairs the total QoC when RXs are distributed at varying distances from their TXs. Whereas, the edge perturbations achieve their best performance in homogeneous networks where the TX-RX distances are consistent across all pairs. In addition, the performance of the proposed adversarial attacks is independent on the SNR values. The results concerning the distribution of eigenvalues indicate that the eigenvalues of $\mathbf{H}$ conform to Johnson's SU distribution before the application of adversarial attacks. Moreover, it is observed that the $B_c$ vertex, $B_p$ vertex, and $B_p$ edge perturbations do not alter the type of distribution of eigenvalues; instead, they only modify the distribution parameters. However, the $B_c$ edge perturbation does change the distribution of eigenvalues. These findings underscore the need for defense mechanisms against adversarial attacks and call for further research on the security and robustness of deep learning-based wireless systems. Future studies may explore adversarial attacks in distributed deployment scenarios.

## APPENDIX
## DISTRIBUTION LIST

The distribution abbreviations used throughout this paper are presented here. Weibull$_{\text{min}}$ and Weibull$_{\text{max}}$ stand for the Weibull Minimum Extreme Value and the Weibull Maximum Extreme Value, respectively. In addition, Gumbel$_r$ and Gumbel$_l$ denote the right-skewed Gumbel and the left-skewed Gumbel, respectively. Moreover, the upper half of a generalized normal and the Reciprocal Inverse Gaussian are represented by HalfGenNorm and Recipinvgauss, respectively.

## REFERENCES

[1] A. Ghasemi, E. Zeraatkar, M. Moradikia, and S. R. Zekavat, "Adversarial attacks on resource management in p2p wireless communications," in *2023 IEEE International Conference on Wireless for Space and Extreme Environments (WiSEE)*, pp. 148–153, 2023.

[2] A. Ghasemi and S. A. Zekavat, "Low-cost mmwave mimo multi-streaming via bi-clustering, graph coloring, and hybrid beamforming," *IEEE Transactions on Wireless Communications*, vol. 20, no. 7, pp. 4113–4127, 2021.

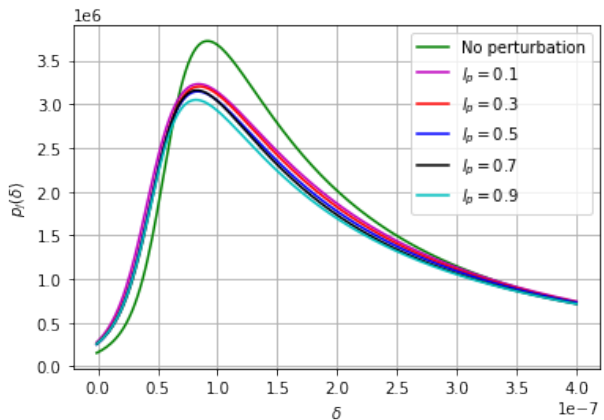

Fig. 12: PDF for EVs after $B_p$ edge perturbation.

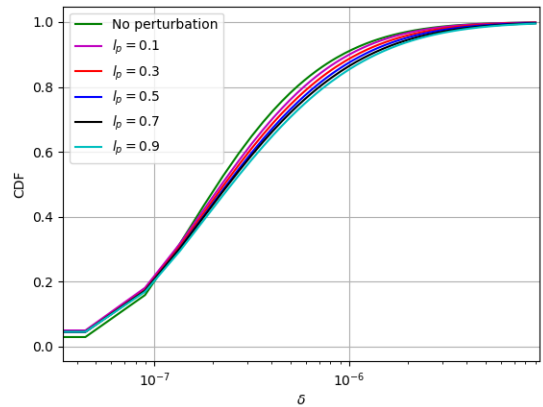

Fig. 13: CDF for EVs after $B_p$ edge perturbation.

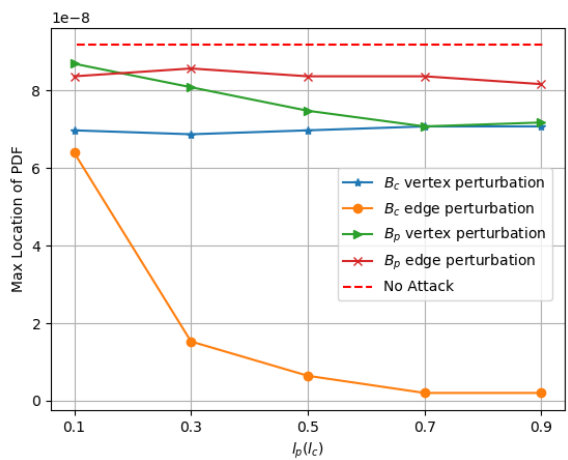

Fig. 14: Locations of the PDF maximum value before and after perturbations.